# Polaronic metal phases in La$_{0.7}$Sr$_{0.3}$MnO$_3$ uncovered by inelastic neutron and x-ray scattering

M. Maschek[1], D. Lamago[1,2], J.-P. Castellan[1,2], A. Bosak[3], D. Reznik[4] & F. Weber[1]

[1] *Institute for Solid State Physics, Karlsruhe Institute of Technology, D-76021 Karlsruhe, Germany*
[2] *Laboratoire Léon Brillouin (CEA-CNRS), CEA-Saclay, F-91911 Gif-sur-Yvette, France*
[3] *European Synchrotron Radiation Facility, F-38043 Grenoble Cedex, France*
[4] *Department of Physics, University of Colorado at Boulder, Boulder, Colorado, 80309, USA*

**Among colossal magnetoresistive manganites the prototypical ferromagnetic manganite La$_{0.7}$Sr$_{0.3}$MnO$_3$ has a relatively small magnetoresistance, and has been long assumed to have only weak electron-lattice coupling. Here we report that La$_{0.7}$Sr$_{0.3}$MnO$_3$ has strong electron-phonon coupling: Our neutron and x-ray scattering experiments show strong softening and broadening of transverse acoustic phonons on heating through the Curie temperature $T_C = 350$ K. Simultaneously, we observe two phases where metallic resistivity and polarons coexist. The ferromagnetic polaronic metal phase between 200 K and $T_C$ is characterized by quasielastic scattering from dynamic CE-type polarons with the relatively short lifetime of $\tau \approx 1$ ps. This scattering is greatly enhanced above $T_C$ in the paramagnetic polaronic metal phase. Our results suggest that the strength of magnetoresistance in manganites scales with the inverse of polaron lifetime, not the strength of electron-phonon coupling.**

In colossal magnetoresistive (CMR) manganites a transition from a ferromagnetic (FM) metallic ground state to a paramagnetic (PM) insulating phase at elevated temperatures increases with an applied magnetic field, which favors the FM phase. In the insulating phase Jahn-Teller (JT) interactions between the carriers and the atomic lattice favor local lattice distortions that trap the charge carriers [1-3]. In many manganites these lattice distortions form superstructures on heating through $T_C$ as a result of CE-type [4,5] short-range charge and orbital order (COO) of Mn$^{3+}$ and Mn$^{4+}$ ions [6-8]. According to Hund's rules, three of the four 3$d$ electrons of the JT active Mn$^{3+}$ ion in La$_{1-x}$Sr$_x$Mn$^{3+}_{1-x}$Mn$^{4+}_x$O$_3$ occupy the lower $t_{2g}$ orbitals forming a $S = 3/2$ core spin, whereas the remaining electron can choose between the two energetically degenerate $e_g$ levels. Such a system will lower the energy of one of the $e_g$ levels by a structural distortion of the Mn$^{3+}$O$_6$ octahedron via the JT effect [9]. For half- doped manganites with an even mix of JT active Mn$^{3+}$ and JT inactive Mn$^{4+}$ ions, Goodenough proposed [5] a CE-type COO ground state with, a crystal lattice distortion with a wave vector of $\mathbf{q}_{CE} = (¼, ¼, 0)$. It is observed in La$_{0.5}$Ca$_{0.5}$MnO$_3$ [10,11], LaSr$_2$Mn$_2$O$_7$ [12], and similar compounds. For smaller doping, the imbalance between Mn$^{3+}$ and Mn$^{4+}$ prohibits long range COO. Here, an $e_g$ electron is able to hop from a Mn$^{3+}$ to a Mn$^{4+}$ ion and the intermediate oxygen atom. Due to the strong Hund's coupling of the 3$d$ electrons the hopping probability is maximized for a ferromagnetic alignment of the $t_{2g}$ Mn core spins. Due to this double exchange (DE) mechanism, ferromagnetism and an undistorted lattice win below $T_C$. Above $T_C$ JT interactions favour short range COO associated with lattice distortions localizing the charge carriers [8,13,14].

Experimentally, there is a clear correlation between the magnitude of magnetoresistance and $T_C$: The resistivity jump increases with decreasing $T_C$. Standard theory correlates this $T_C$ reduction with increasing electron-phonon coupling (EPC) strength [1,2,15,16]. In particular, models of CMR assume that the amplitude of the lattice distortions controls how much the resistivity increases above the Curie temperature, T$_C$.

Among ferromagnetic manganites, La$_{0.7}$Sr$_{0.3}$MnO$_3$ (LSMO) has the highest $T_C = 350$ K, and a comparatively small CMR effect [17]. LSMO undergoes a metal-metal transition at $T_C$ as opposed to the metal-insulator transition in, e.g. La$_{0.7}$Ca$_{0.3}$MnO$_3$ [18] and La$_{1.2}$Sr$_{1.8}$Mn$_2$O$_7$ [19]. So the point of view emerged that the metallicity, high T$_C$, and small CMR effect in LSMO, are well explained by DE physics without JT polarons [3]. However, in a recent experiment [20] some of us demonstrated that the lattice distortions in LSMO increase at $T_C$ similarly to other manganites having a much larger magnetoresistance effect. Small polarons in LSMO were reported by pulsed neutron diffraction as well [a] [21]. These experiments did not have energy resolution, so no information was obtained on electron-phonon effects for specific phonons nor on the spacial or temporal character of the polarons.

These results motivated us to investigate lattice dynamics of LSMO in more detail. Using direct measurements of phonons and polarons by energy-resolved neutron and x-ray scattering, we report both strong electron-phonon coupling for phonons near $\mathbf{q}_{CE}$ and formation of dynamic polarons of CE-type in the ferromagnetic as well as paramagnetic phases.

---

[a] We will show in the following that the results of Chen et al. [26] are signatures of the temperature dependent renormalization of the acoustic phonon mode and do not, as suggested by the authors, reveal the existence of polaronic fluctuations at finite energy transfers.

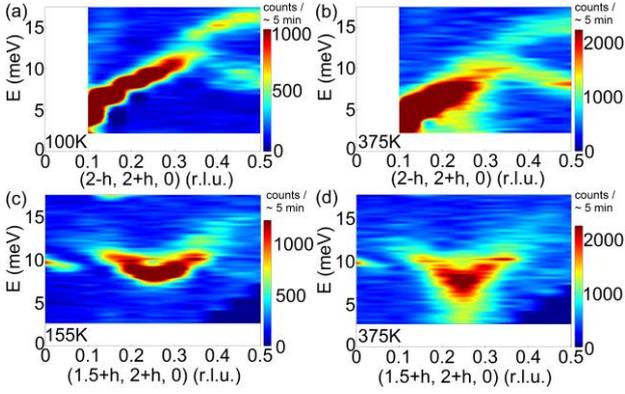

**Figure 1** (color online): *(a),(b)* Color-coded contour plots of raw neutron scattering data obtained for $La_{0.7}Sr_{0.3}MnO_3$ on the thermal TAS 1T, LLB, at $T = 100$ K and $T = 375$ K. Constant $Q$ scans at $Q = (2-h, 2+h, 0)$, $0.1 \leq h \leq 0.5$ and $\Delta h = 0.05$ r.l.u., were performed with energy steps of $\Delta E = 0.25$ meV. For readability, the intensity scale was adjusted at different temperatures according to the Bose factor of a phonon at an energy of $E = 9$ meV. *(c)(d)* Color-coded contour plots of raw neutron scattering taken at $T = 155$ K and $T = 375$ K along an off-symmetry direction $Q = (1.5+h, 2+h, 0)$, $0 \leq h \leq 0.5$, $\Delta h = 0.05$ r.l.u. and $\Delta E = 0.25$ meV.

Inelastic neutron and x-ray scattering measurements were performed at the 1T neutron triple-axis spectrometer located at LLB (CEA Saclay) and the high energy resolution x-ray spectrometer ID-28 at ESRF (Grenoble), respectively. In both setups we focused on the transverse acoustic (TA) phonon mode propagating along the [110] axis, i.e. wave vectors $Q = (2-h, 2+h, 0)$ where wave vectors are given in the pseudo-cubic perovskite structure of LSMO in units of $(2\pi/a, 2\pi/a, 2\pi/a)$ with $a = 3.88$ Å. Our sample was a high-quality single crystal of $La_{0.7}Sr_{0.3}MnO_3$ weighing 4 g already used in previous experiments [22]. The sample for x-ray scattering was prepared from a piece cut from the large sample after the neutron scattering experiments were finished. The experimental resolution in the neutron scattering experiments was obtained from standard calculations using the "*rescal*" program package [23]. In x-ray scattering, the resolution was determined by scanning the elastic line of a piece of plastic at the beginning of the experiment.

Our experiments mapped reciprocal space in the vicinity of the $\Sigma_3$ TA phonons dispersing in the [110] direction at different temperatures, $T$. These phonons have the same symmetry as the JT distortion of the $MnO_6$ octahedra [14,24,25] and that of the charge order peaks observed in the insulating paramagnetic phases of CMR manganites such as $La_{1.2}Sr_{1.8}Mn_2O_7$ [13] or $La_{0.7}Ca_{0.3}MnO_3$ [8]. The structure factor of these phonons is large next to the $\tau = (2,2,0)$ reciprocal lattice vector, where it can be measured in a purely transverse geometry, i.e. $\tau \cdot q = 0$ at $Q = \tau + q = (2,2,0) + (-h, +h, 0)$. To compare our results more closely with published results [26] we also studied scattering intensities at $Q = (1.5+h, 2+h, 0)$, i.e. perpendicular to the direction discussed above and intersecting at $h = 0.25$.

At low temperatures our measurements reveal well defined phonon dispersions and sharp peaks [Figs. 1(a)(c)]. The TA branch at $q = (-h, h, 0)$ can be easily followed up to $h = 0.3$ where the intensity is dominated by a single peak[b] [Fig. 1(a)]. Near the zone boundary, i.e. $h \geq 0.45$, we observe two well separated peaks at 8.5 meV and 16.25 meV. Close to $h = 0.35$ raw data show an anti-crossing with a transverse optic (TO) branch in agreement with previous measurements [24]. This happens when upward and downward dispersing phonon branches of the same symmetry come close to each other. They are not allowed to cross but can exchange eigenvectors. Hence, close to the zone boundary the TA character is transferred to the nominal first TO branch, i.e. the TA character is mostly in the peak at 16.25 meV.

Our data along the $Q = (1.5+h, 2+h, 0)$ direction at $T = 155$ K [Fig. 1(c)] also clearly show the TA mode. As we leave the transverse geometry with increasing $|h - 0.25|$, the TA phonon structure factor quickly decreases and we lose the peak for $h < 0.1$ and $h > 0.4$. At $T = 375$ K, the TA phonon at $q_{CE} = (¼, ¼, 0)$ acquires a pronounced low energy tail, whereas the phonons at other wave vectors, e.g. at the zone boundary ($h = 0.5$), do not strongly depend on temperature [Fig. 1 (b)(d)].

The constant-momentum-transfer scans were fitted with damped-harmonic-oscillator (DHO) functions convoluted with the calculated Gaussian experimental resolution [27]. The experimental background for each temperature was taken from the scans at the zone boundary [Figs. 2(a)(b)] for $E \leq 6$ meV, and near the zone center for $E \geq 12$ meV and approximated by a straight line. The resulting fits are shown in Fig. 2 for $h = 0.5$ [Figs. 2(a)(b)] and 0.25 [Figs. 2(c)(d)]. At $T = 200$ K (Fig. 2, left column) the phonon line widths are close to the calculated resolution and the experimental background is reached on either side of the observed modes.

At $h = 0.5$ (the zone boundary) the phonon softens by 3% on heating to 375 K [Fig. 3(c)] as expected from thermal expansion and atomic motions leading to softer bonds. Its intensity follows the Bose factor. At $h = 0.25$ [Figs. 2(c),(d)] the temperature dependence is clearly unconventional in that the phonon softens by more than 10% [Fig. 3(c)] and broadens substantially [Fig. 3(d)] and neutron counts do not reach the background at low energies at high temperatures.

---

[b] In our scans we see also a small signal around 8 meV ($h = 0.15$) and 10 meV ($h = 0.25$), which we ascribe to spurious scattering. It is most pronounced at low temperatures and harder to observe when the phonon broadens and thereby masks the spurious scattering. As the signal intensity is small compared to that of the TA mode and does not change with temperature it did not affect our analysis.

The temperature dependence of the integrated phonon intensity is set by the Bose factor built into the DHO fits, which indicate that only half of the scattering intensity at $E = 2 - 3$ meV for $h = 0.25$ and $T = 375$ K [Fig. 2(d)] can be attributed to the TA phonon. Put differently, phonon line broadening cannot explain all scattering intensity at low energies at $h = 0.25$. The fits to the data can be much improved by introducing a Lorentzian centered at zero energy indicating the presence of quasielastic (QE) scattering. Unfortunately, the neutron scattering data are not sufficient to independently extract the QE scattering amplitude and line width because we could not measure below an energy transfer of 2 meV for technical reasons.

To further investigate this QE scattering, we performed high-energy resolution inelastic x-ray scattering (IXS), which has an effectively unlimited energy range. Above 200 K we see substantial QE intensity at $\mathbf{Q} = (1.75, 2.25, 0)$ [Figs. 2(f)] in addition to the Bose factor already included in the DHO fits, which can be well fitted by a Lorentzian line. Data taken at $\mathbf{Q} = (2-h, 2+h, 0)$, $100\,\text{K} \leq T \leq 400\,\text{K}$ and $-22\,\text{meV} \leq E \leq +22\,\text{meV}$ were analyzed in the same way as the neutron results, except that measuring negative energies in IXS allowed us to make a free fit of the Lorentzian QE line.

Figure 3(a)-(d) shows that the QE intensity increases with temperature above 200 K and peaks at $T = 350$ K [Figs. 2(f) and 3(b)], which is the Curie temperature $T_C$ determined by magnetic neutron diffraction[c] [Fig. 3(a)]. After removing the effect of the resolution, the energy width of the fitted Lorentzian at 350 K is 4.15 meV (half width at half maximum, *HWHM*), which corresponds to the lifetime of $t = (1.0 \pm 0.15)\,\text{ps}$. It does not vary significantly with $T$ compared to the uncertainty. Hence, we used this line width value also to analyze the neutron data, from which it cannot be determined independently.

The resulting phonon energies and line widths of the TA mode at $h = 0.25$ show strong softening [Fig. 3(c)] and broadening [Fig. 3(d)]. The conventional effects of thermal expansion and thermal induced disorder are much smaller as can be seen in the results for the TA zone boundary mode at $h = 0.5$.

Moreover, the observed wave vector dependences show that both QE scattering and phonon renormalization are strongest close to $\mathbf{q}_{CE}$ [Figs. 3(e)-(h)]. Indeed, the wave vector range of anomalous effects along the [110] direction is the same for phonon and quasielastic scattering and we can extract a correlation length for the quasielastic scattering of $\xi = (34 \pm 4)\,\text{Å}$. Hence, the effects are both clearly linked to CE-type [4,5] short-range COO fluctuations.

Here, we want to comment on a previous neutron scattering report on dynamic correlated polarons in

---

[c] Because $(1,0,0)$ Bragg peak measurements were performed with an imperfect energy resolution, the measured intensity includes magnetic fluctuations. Thus we assigned $T_C$ to zero of the second derivative of the curve in Fig. 3(a).

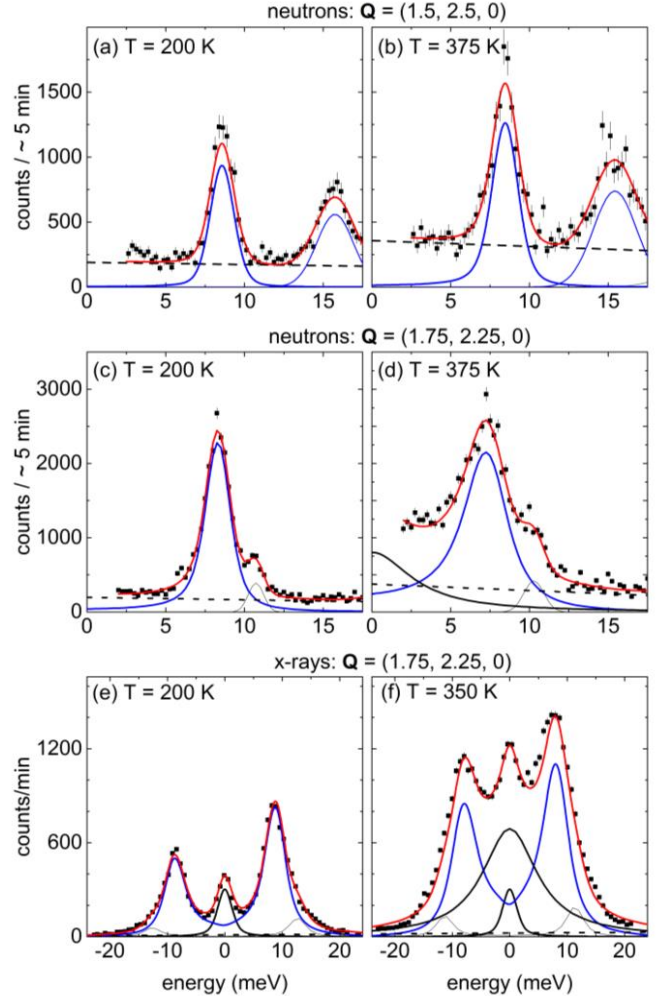

**Figure 2** (color online): Representative inelastic neutron *(a)-(d)* and x-ray *(e)-(f)* scattering spectra of La$_{0.7}$Sr$_{0.3}$MnO$_3$ well below *(a),(c),(e)* and just above *(b),(d),(f)* the Curie temperature $T_C = 350$ K. Results from neutron scattering are shown for $Q = (2-h, 2+h, 0)$ with $h = 0.5$ (top row) and $h = 0.25$ (middle row). X-ray scattering data are presented for $h = 0.25$ (bottom row). Solid (red) lines are fits consisting of a damped harmonic oscillator function for the TA mode (solid & blue), a Lorentzian for quasielastic scattering described in the text (solid & black) and the estimated background (dashed, see text). The small Gaussian component (thin black line) in *(c)-(f)* denotes a spurious signal (see text). In x-ray scattering we also observe a practically temperature independent incoherent elastic line, which is resolution limited.

LSMO [26]. This study showed that constant energy scans at $E = 6$ meV along $\mathbf{Q} = (1.5+h, 2+h, 0)$ have a peak at $h = 0.25$ appearing above $T = 200\,K$. This is consistent with our data [Figs. 1(c)(d)]: At low temperatures such a scan probes only the (flat) background, whereas the softening of the TA phonon leads to increasing intensities near $h = 0.25$ at $T > 200\,K$. Our detailed analysis [Figs. 2(c),(d)] shows, however, that this increase at $E = 6$ meV is largely due to the softening of the TA phonon.

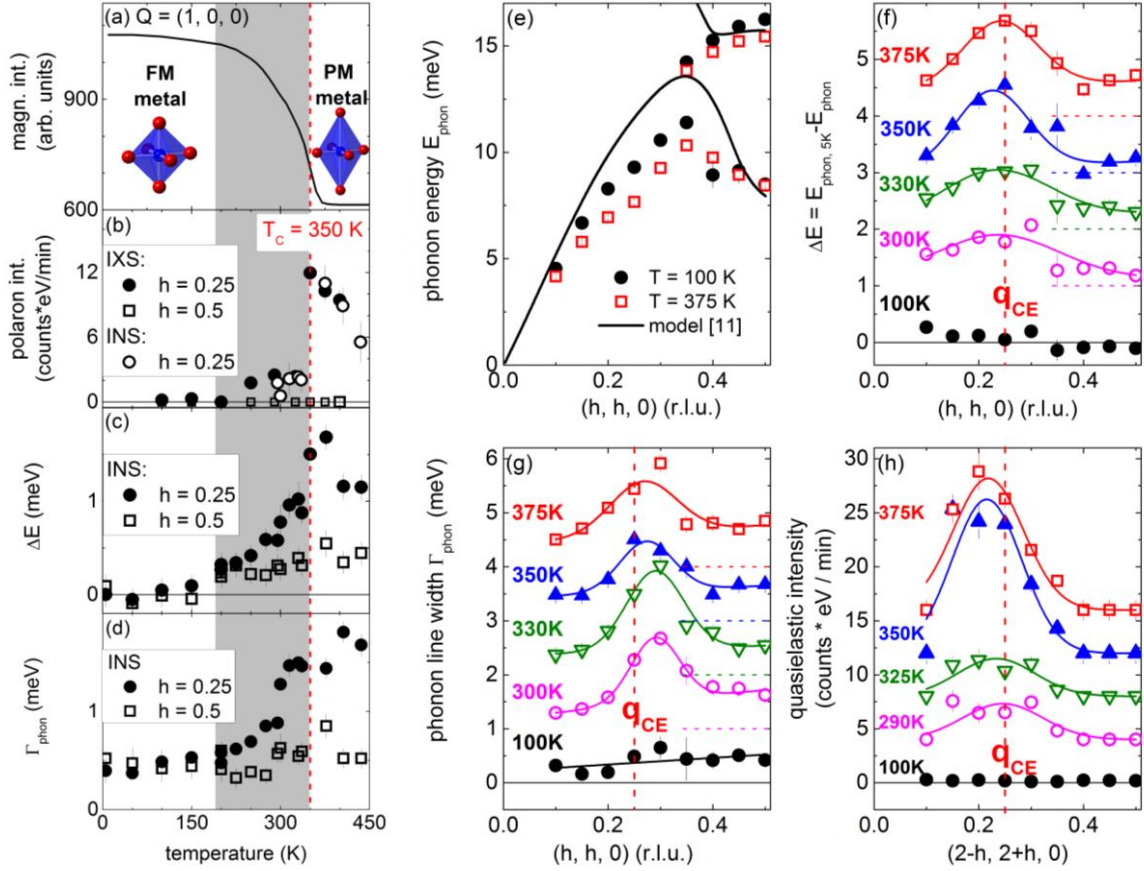

**Figure 3** (color online): Wave vector and temperature dependences of phonon renormalization and quasielastic scattering in $La_{0.7}Sr_{0.3}MnO_3$. Temperature dependences of *(a)* the ferromagnetic intensity of the (100) Bragg position and *(b)* the quasielastic scattering intensity, *(c)* $\Delta E = E_{phon,5K} - E_{phon}(T)$, *(d)* $\Gamma_{phon}$ at $h = 0.25$ (dots) and 0.5 (open squares). INS results for the quasielastic intensity in *(b)* (circles) are scaled to match the high temperature x-ray results (dots). *(e)* Dispersion of the TA phonon along the [110] direction at 100 K and 375 K obtained from inelastic neutron scattering. Lines represent a shell model calculation for a cubic unit cell of $La_{0.7}Sr_{0.3}MnO_3$ reproduced from Ref. [24]. *(f)-(h)* Wave vector dependences at temperatures 100 K $\leq T \leq$ 375 K of *(f)* the observed phonon softening $\Delta E(q)$, *(g)* the phonon line width $\Gamma_{phon}(q)$ (HWHM) and *(h)* the quasielastic intensities associated with CE-type polarons. As the character of the TA phonon is transferred to the nominal first TO near the zone boundary, we plot results of the latter for $0.4 \leq h \leq 0.5$. Data in *(h)* and at $T = 350$ K in *(f)* and *(g)* were obtained by IXS. All other results are based on INS. Solid lines are Gaussian fits to the data. Data are offset for clarity and the corresponding zeros are shown as color-coded dashed lines on the right side of the panel. Vertical dashed lines denote the position of the ordering wave vector of the CE-type COO $\boldsymbol{q}_{CE} = \left(\frac{1}{4}, \frac{1}{4}, 0\right)$.

Strong CE-type polarons that we found in LSMO, are not as obvious as in $La_{1.2}Sr_{1.8}Mn_2O_7$ or $La_{0.7}Ca_{0.3}MnO_3$ because they are broad in energy ($\approx 4$ meV, HWHM) and thus have a lower peak intensity. However, the energy-integrated intensity of polaron scattering in LSMO at $T = T_C$ appears to have the same order of magnitude as in $La_{1.2}Sr_{1.8}Mn_2O_7$ at $T = 128$ K ($T_C = 115$ K), which we studied in a previous investigation [14]. A detailed investigation of how integrated intensity of polaron scattering varies between different manganites is outside the scope of this work and will be performed in the future.

A key difference between $La_{1.2}Sr_{1.8}Mn_2O_7$ and LSMO is that the former has a polaron-metallic ground state[28] characterized by an optic phonon anomaly already at $T = 10$ K [14], whereas LSMO shows no measurable electron-lattice effects at low temperature at either low or high energies [22]. The polaronic metal character of $La_{0.7}Ca_{0.3}MnO_3$ at low temperatures has not been established or ruled out and is a subject of future work.

LSMO shows signatures of a polaronic metal only above 200 K [Figs. 5(e)-(g)]. Above $T_C$ $La_{1.2}Sr_{1.8}Mn_2O_7$ and $La_{0.7}Ca_{0.3}MnO_3$ become paramagnetic polaronic insulators with *static* polarons [8,14], whereas LSMO becomes a paramagnetic polaronic metal with *mobile* polarons. This explains why the absolute strength of magnetoresistance near $T_C$ in LSMO and $La_{1.2}Sr_{1.8}Mn_2O_7$ differs by more than an order of magnitude [20].

Phase separation between charge-ordered (CO) and ferromagnetic metallic regions is discussed in CMR

manganites since more than 15 years [29] and, recently, CMR in undoped $LaMnO_3$ under an applied pressure of $p = 32 - 35$ GPa was also explained in this scenario [30]. In LSMO, the sharp increase of the QE intensity at $T_C$ would indicate the breakdown of the percolation path of ferromagnetic metallic domains and the sudden increase of CO ones while some of the latter already form shortly above $T = 200$ K.

To conclude, our results suggest that strong electron-phonon coupling and polarons are a generic feature of FM manganites. Our results imply the following phase diagram for LSMO: The ground state is a FM-DE metal. For $200\,K \leq T \leq 350\,K$ ($\approx T_C = 370\,K$) LSMO is a FM polaronic metal, whose polaronic character becomes strongly enhanced in the PM polaronic metal phase above $T_C$. This reconciles the need for polarons to explain CMR quantitatively in LSMO at $T \approx T_C$ [1,2] with the observation of a DE metallic ground state, which has a much higher conductivity [17] than the metallic phase in the compounds with strong CMR. Our results suggest that the strength of magnetoresistance in manganites crucially depends on the polaron mobility. Thus, a realistic model of polaron dynamics is essential to understand CMR theoretically. We hope that our work will stimulate development of such theories.

**Acknowledgements:**
M.M. and F.W. were supported by the Helmholtz Society under contract VH-NG-840. D.R. was supported by the DOE, Office of Basic Energy Sciences, Office of Science, under Contract No. DE-SC0006939.